\documentclass[12pt,halfline,letterpaper]{ouparticle}
\usepackage{listings}
\usepackage{hanging}
\usepackage{censor}
\usepackage[font=small,labelfont=bf]{caption}
\usepackage{color, colortbl}
\definecolor{Gray}{gray}{0.9}

\begin{document}

\title{National Power as Network Flow}

\author{%
\name{Michael Poulshock}
}

\abstract{Political power in the international context can be characterized as a fluid-like substance that circulates through a network of nation states. States can possess it as a stock quantity, reflected by their material capacity or national wealth; and they can transfer it as a flow quantity, through constructive or destructive action. Constructive activities like trade increase a state's power, while destructive ones like violent conflict reduce it. In this paper, we quantify these assertions to a first approximation using economic and military data, parameterizing a mathematical model that can forecast the evolution of power in the international system.}

\maketitle

\section{Introduction}

National power seems like something that should be quantifiable. It is obvious that some states have more of it than others, that these power levels vary over time, and that states through various activities try to maximize or at least maintain the amount of power that they possess. Since the 1940s, numerous efforts have been made to devise quantitative measurements of national power (Tellis 2000). These metrics typically combine variables related to the economic, military, demographic, and geographic characteristics of a country into a single number to facilitate inter-country comparisons. Despite the profusion of definitions that have been developed, most produce similar rank order results (Tellis, ch. 3, citing Merritt and Zinnes).

In this paper, we propose a new framework for quantifying national power. It too produces rank order results similar to existing measures. However, unlike prior efforts, it also describes how power levels change over time as a result of the positive and negative interactions among states. The essential idea is to model the international system as a \textit{power structure}, an object composed of states of varying capacities, or \textit{sizes}, along with their relationships (Poulshock 2019). These relationships can be positive (representing constructive action), negative (representing destruction), or neutral. Power structures abstract international relations down to its most basic elements: nations of different strengths interacting with each other. Power structures can be readily visualized as graphs, such as in Fig. 1, in which larger nodes are more powerful, solid lines indicate cooperation, and dashed lines indicate conflict.

\begin{figure}[H]
\centering
\includegraphics[scale=.25]{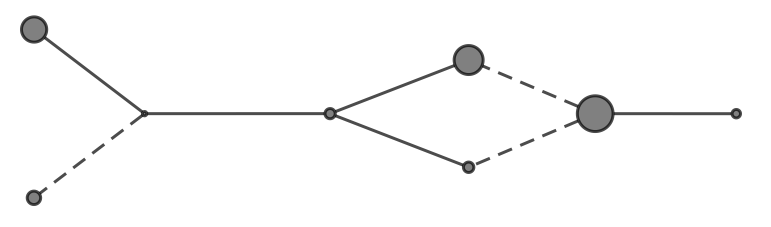}
\caption{An abstract power structure.}
\end{figure}

\textit{Power} is defined as the ability to affect another state's power, either positively or negatively. The axioms underlying this conception of power are: (1) when states cooperate, they get stronger; (2) when states fight, they weaken each other; and (3) unused power depreciates relative to power used constructively (Poulshock 2019). Applying these principles, we can envision political power as a kind of fluid that sloshes around the network, causing some states to grow and others to shrink as a result of their relationships. This \textit{law of motion} is illustrated in Fig. 2. Power is therefore both a stock, possessed by states, and a flow, in motion from one state to another.

\begin{figure}[H]
\centering
\includegraphics[scale=.25]{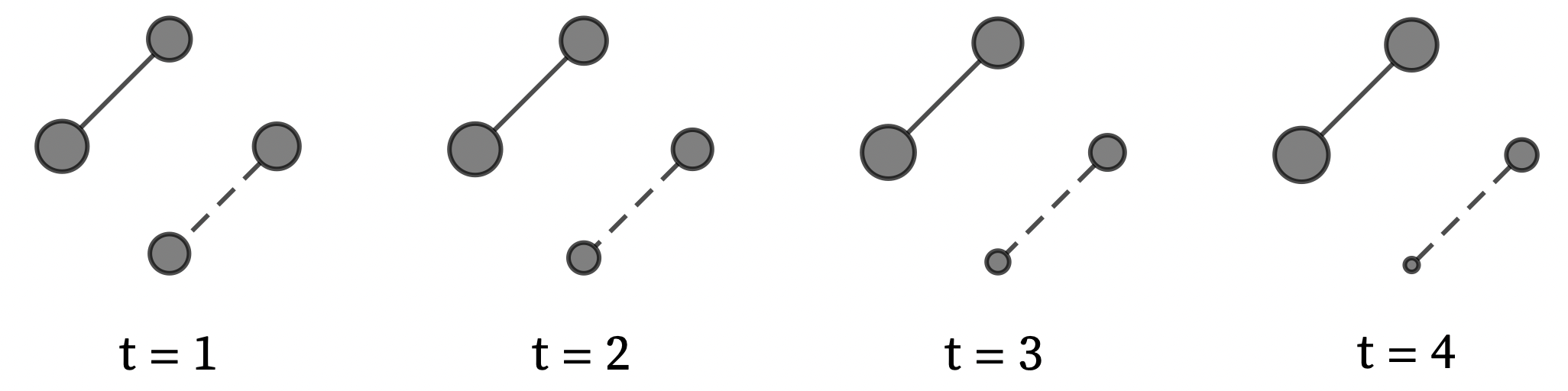}
\hspace{3em}
\includegraphics[scale=.2]{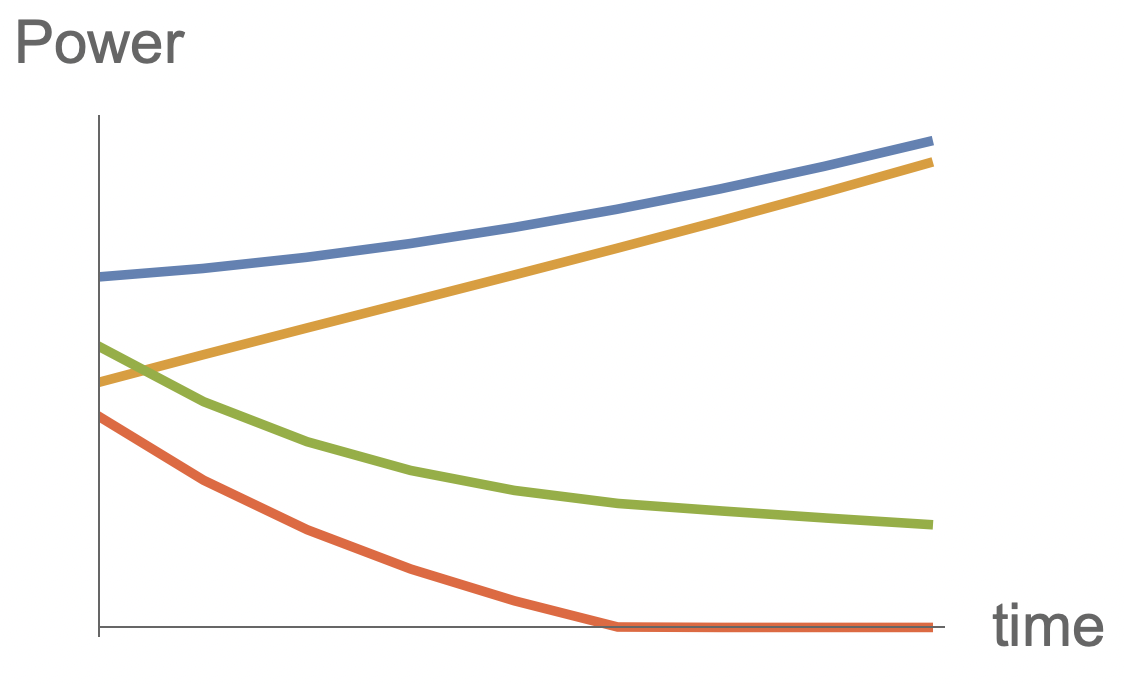}
\caption{Two ways of illustrating the law of motion.}
\end{figure}

Our objective is to apply this model to the international system. A threshold question is which real world quantities best correspond to power in the abstract model. A number of considerations restrict this choice: (1) the variable corresponding to power at rest must be a stock variable; (2) the variables corresponding to constructive and destructive power must be flow variables; (3) power at rest must be convertible into constructive and destructive flows; (4) destructive flows must tend to decrease power at rest; (5) constructive flows must tend to increase power at rest; (6) since power at rest is interpreted broadly as the capacity to impose consequences on another state, its real world analogue should encompass the widest possible spectrum of activity; (7) the quantity for power at rest must allow for \textit{absolute} increases and decreases in power;\footnote{This follows from axioms 1 and 2. Many definitions of national power are based on states' shares of the world total of some quantity, such as GDP, and therefore do not meet this criterion.} and (8) quantities that are combined together should have compatible units of measure.\footnote{Many definitions of national power obviate this requirement by adding together shares of world totals of otherwise incompatible quantities, such as population and GDP.}

Given these constraints, we contend that national power is naturally, perhaps inevitably, estimated by national wealth. National wealth reflects the total material assets of a nation. It is a stock quantity,\footnote{GDP is a flow rather than a stock quantity, and therefore not a good candidate for power at rest.} estimated at a point in time using a standardized accounting definition (United Nations System of National Accounts (2008), \S\,13.4). It is reasonably in accord with our intuitions about national strength as material capacity, and likely to produce rank orderings similar to other proposed definitions of national power (see Fig. 3).

\begin{figure}[H]
\centering
\includegraphics[scale=.35]{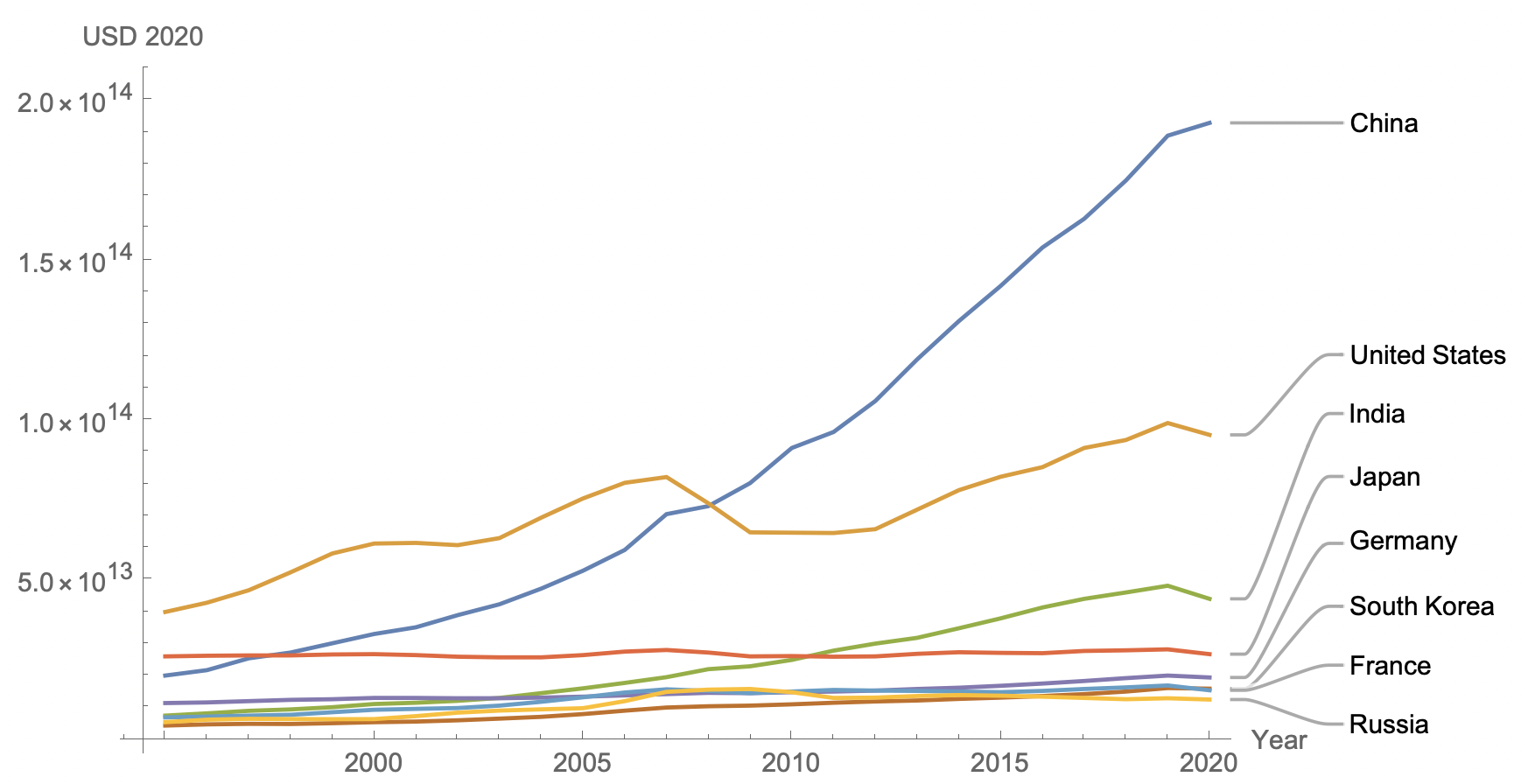}
\caption{Largest countries by national wealth (1995-2020).}
\end{figure}

The flow variables for destructive and constructive actions can be quantified consistently with the use of national wealth as the stock variable. Destructive action can be measured by state A's military expenditure on a particular conflict with state B over some time interval, along with a concomitant reduction in state B's wealth due to the effects of applied violence. In most conflicts, these transactions are bidirectional, with both parties harming each other. Each transaction has two components: the amount expended by the aggressor and the amount of damage suffered by the recipient. The expenditure would ideally be measured by the value of the aggressor's military assets before the conflict minus their value after. The loss of power that results from being a recipient of violence would be quantified by the reduction in assets due to destruction. Note that this latter relationship is also an accounting identity: the destruction of assets due to violent conflict is recorded as a catastrophic loss in the System of National Accounts.\footnote{Catastrophic losses roll up into ``other changes in the volume of assets," per section 12.46 of (United Nations 2008).} So in this sense it is true \textit{by definition} that war reduces national wealth.

Constructive action can in turn be approximated by the volume of commerce between two countries. For the most part, this can be estimated as the total imports and exports of goods and services, including government-to-government transfers. In an economic transaction, both buyer and seller give something up but receive something else of greater value to them; otherwise, they would not trade. The model assumes that this mutual benefit ultimately contributes to an increase in national wealth. 

All of these variables relate to each other conceptually and can be expressed in terms of a reference currency. For these reasons, we contend that the appropriate way to estimate national power as a stock quantity is to use national wealth, and that power flows can be estimated using international trade volumes and changes in military asset levels. 

\begin{figure}[H]
\small
\centering
\begin{tabular}{ | l | l | c | }
\hline
\rowcolor{Gray}
\textbf{Manifestation} & \textbf{Variable} & \textbf{Type} \\
\hline
Power at rest & National wealth & Stock \\ \hline
Constructive power & International trade & Flow \\ \hline
Destructive power & Changes in military assets & Flow \\
\hline
\end{tabular}
\caption{Manifestations of power and their corresponding proxy variables.}
\end{figure}

In this paper, we parameterize the power structure model using real world wealth, trade, and military data from 193 countries over the period 1995-2020. We then illustrate how it can be used to simulate the evolution of power in the international system.

In applying this model to international relations, we are not asserting that wealth and power are equivalent. In the abstract model, constructive and destructive stocks of power are fungible and can be immediately converted into one another. In reality, it takes time to convert wealth into military assets and it is possible for a wealthy country to have few such assets. So a state can be wealthy but not \textit{powerful} in the traditional sense of being able to effectuate destruction. Ideally, to completely quantify power, we would be able to account for the value of both military and non-military assets and track how they change as a result of the constructive and destructive interactions among states. However, that would require data that is not presently available as well as a more complicated law of motion. Nonetheless, we can still demonstrate the validity of the simplified law of motion as defined in this paper. We just need to bear in mind that national wealth is a proxy variable for power that does not necessarily convey information about the destructive capacity of each state.\footnote{Though we use this model to predict power as approximated by wealth, this is not a theory of national wealth. There already exists a body of economic ideas and a sophisticated accounting framework relating wealth to production, exchange, consumption, and depreciation (Piketty 2014).}

\section{Methodology}

\subsection{Model Calibration}

In order to apply the model, we need to estimate values for three parameters, specifically: To what degree does constructive action increase another state's power ($\beta$)? To what degree does destructive action decrease another state's power ($\mu$)? What happens to a state's power when it has no relations with other states ($\lambda$)? These parameters quantify the three axioms stated above. We do not seek answers that hold across all historical, geographical, and technological circumstances; we merely want a first approximation reflective of the contemporary international system. Our procedures for estimating these parameters and the data sources used are described in Appendix B. Here we summarize the methodology and the resulting parameters.\footnote{All of  the code and data used in this paper is available at https://github.com/mpoulshock/National PowerAsNetworkFlow.}

The model has two parameters that establish growth\footnote{When we refer to \textit{growth}, we mean the increase in power quantified as national wealth, not the rate of productivity (i.e. GDP) growth.} rates for power. The first, $\beta$, quantifies the magnitude of increase when constructive power is transferred from one state to another. The second, $\lambda$, defines the \textit{intrinsic growth rate} for power that is not allocated to other states, in other words, kept as a stock. There can be a wide variation in the intrinsic growth rate from one country to the next, and from one year to the next, due to exogenous variables like productive capacity, internal conflict, changes in supply and demand, the vagaries of financial markets, and the effect of natural disasters. However, to keep things simple, in this paper we use a global value for $\lambda$. To estimate it, we obtained a linear relationship between growth and trade, suggesting that an increase of 2.5\% is intrinsic. We then ran simulations to find the best fit for $\beta$. We estimate that $\beta = 1.392$ and $\lambda = 1.025$.

The parameter $\mu$ quantifies the degree to which destructive action decreases another state’s power (that is, if a state expends $x$ units of power on destructive action, the recipient's stock of power decreases by $\mu x$). To estimate $\mu$, we looked at civil wars and compared those countries' annual military expenditures with the resulting reduction in national wealth. Though far from perfect, this methodology suggests that $\mu$ is in the neighborhood of 30.

\begin{figure}[H]
\small
\centering
\begin{tabular}{ | c | l | c | c | c | }
\hline
\rowcolor{Gray}
\textbf{Parameter} & \textbf{Meaning} & \textbf{Axiom} & \textbf{Estimated Value}\\
\hline
$\beta$ & Effect of constructive action & 1 & 1.392 \\ \hline
$\mu$ & Effect of destructive action & 2 & 30 \\ \hline
$\lambda$ & Intrinsic growth rate & 3 & 1.025 \\ \hline
\end{tabular}
\caption{Summary of the model parameters. }
\end{figure}

\subsection{Simulation}

Having calibrated the model, we can simulate the evolution of power in the international system. To initialize a simulation, wealth, trade, and military data from a given year is loaded into the data structures and parameters described in Appendix A. The power of each country, quantified as national wealth, is then estimated by applying the law of motion iteratively for the desired number of years.

We use two types of simulations. In the first type, which we call \textit{naive}, we apply the law of motion iteratively to the wealth, trade, and conflict statistics from a single year. The word naive refers to the fact that this type of simulation assumes that none of the states change their trade and military relations from one year to the next, an obviously tenuous assumption. The second type, which we refer to as \textit{dynamic}, is initialized with wealth data from a starting year, but the trade and military data changes from year to year over the course of the simulation.

\section{Results and Discussion}

\subsection{Backtesting}

The application of the model can be illustrated using dynamic simulations that predict a country's wealth based on trade and military time series data. Figure 6 shows four such simulations. Strictly speaking, they do not validate the model because they are based on the same data that was used to determine the parameters. Nonetheless, these backtests show that the model is capable of reflecting general trends in the evolution of power.

\begin{figure}[H]
\centering
\includegraphics[scale=.23]{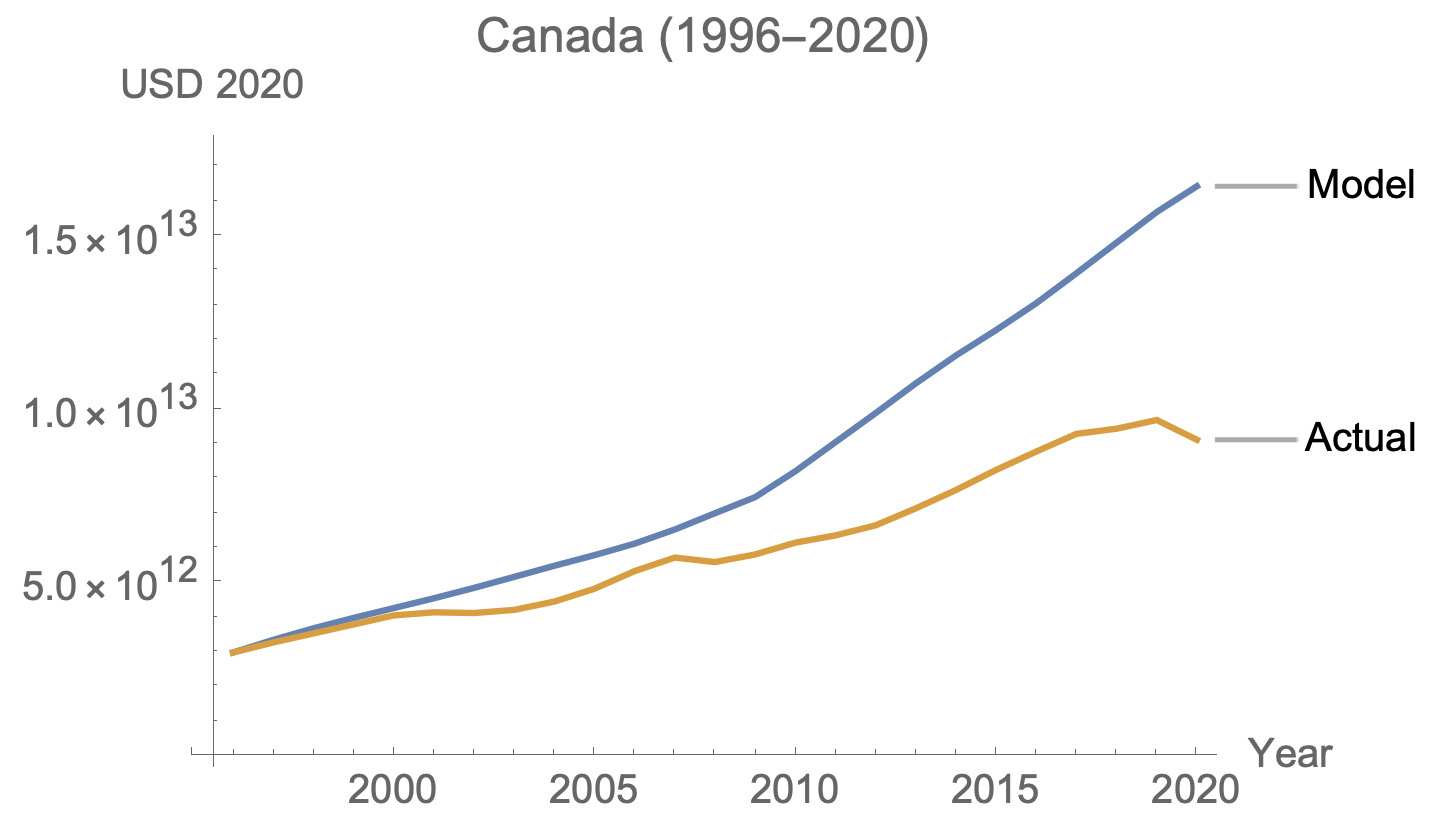}
\hspace{2em}
\includegraphics[scale=.23]{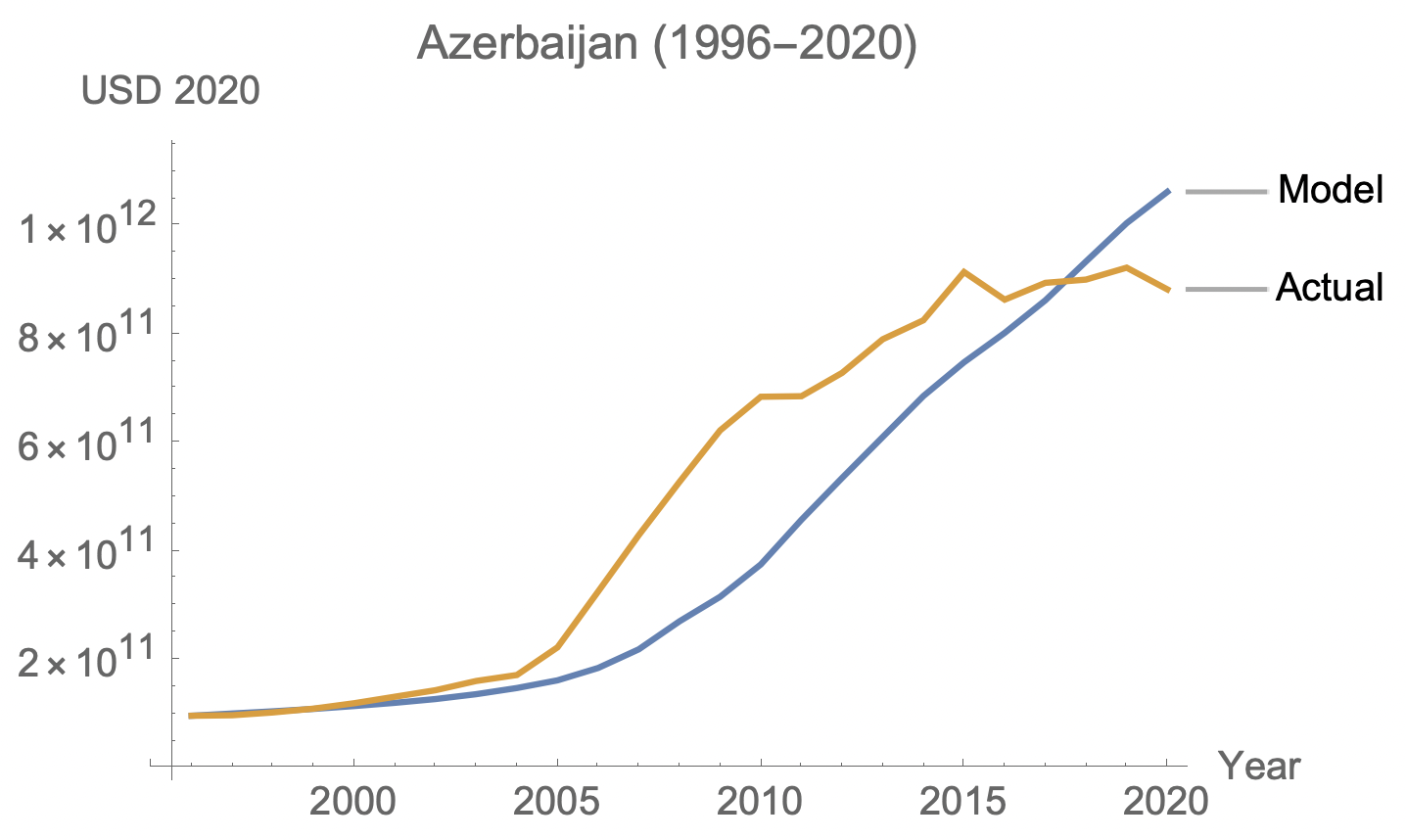}
\includegraphics[scale=.23]{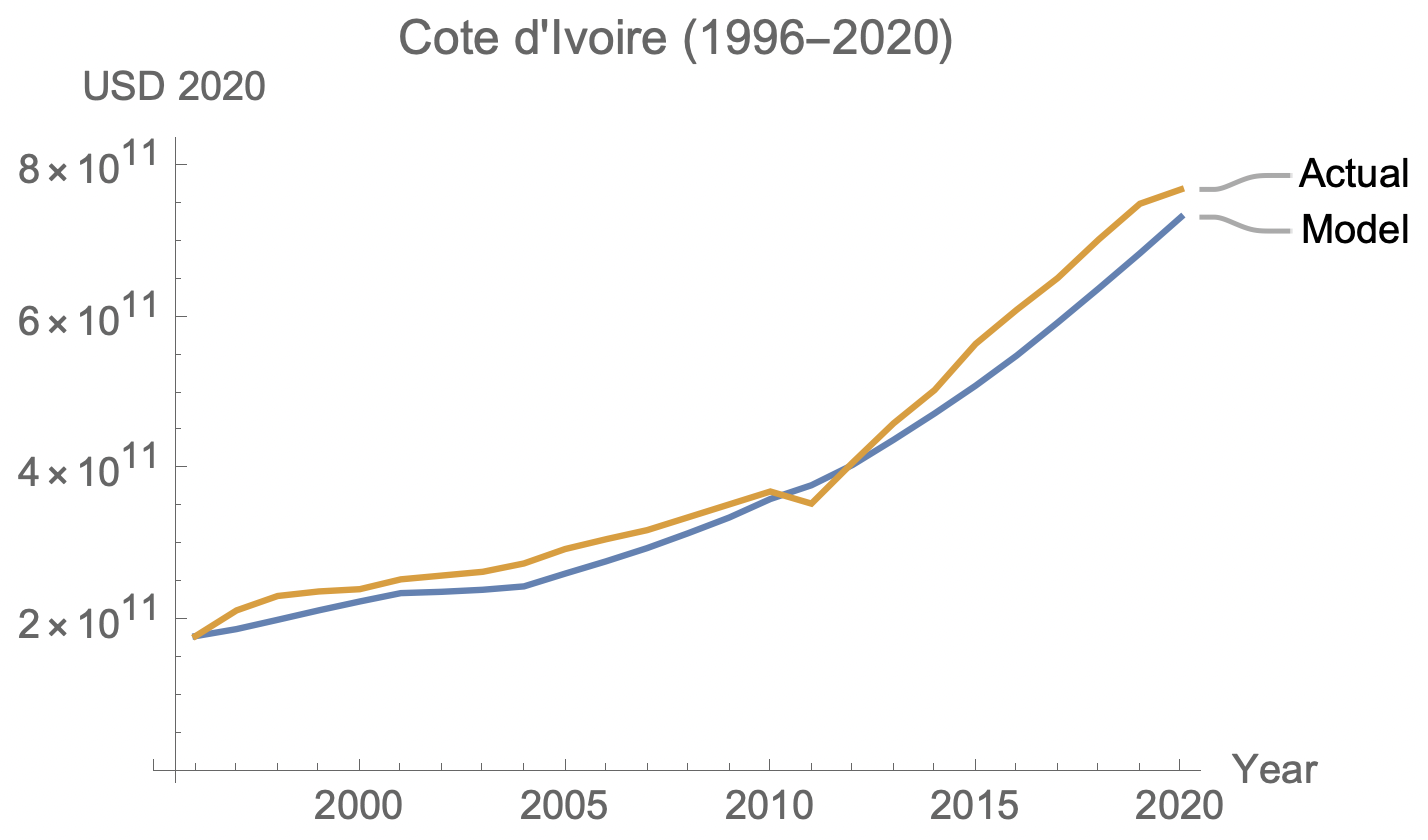}
\hspace{1em}
\includegraphics[scale=.23]{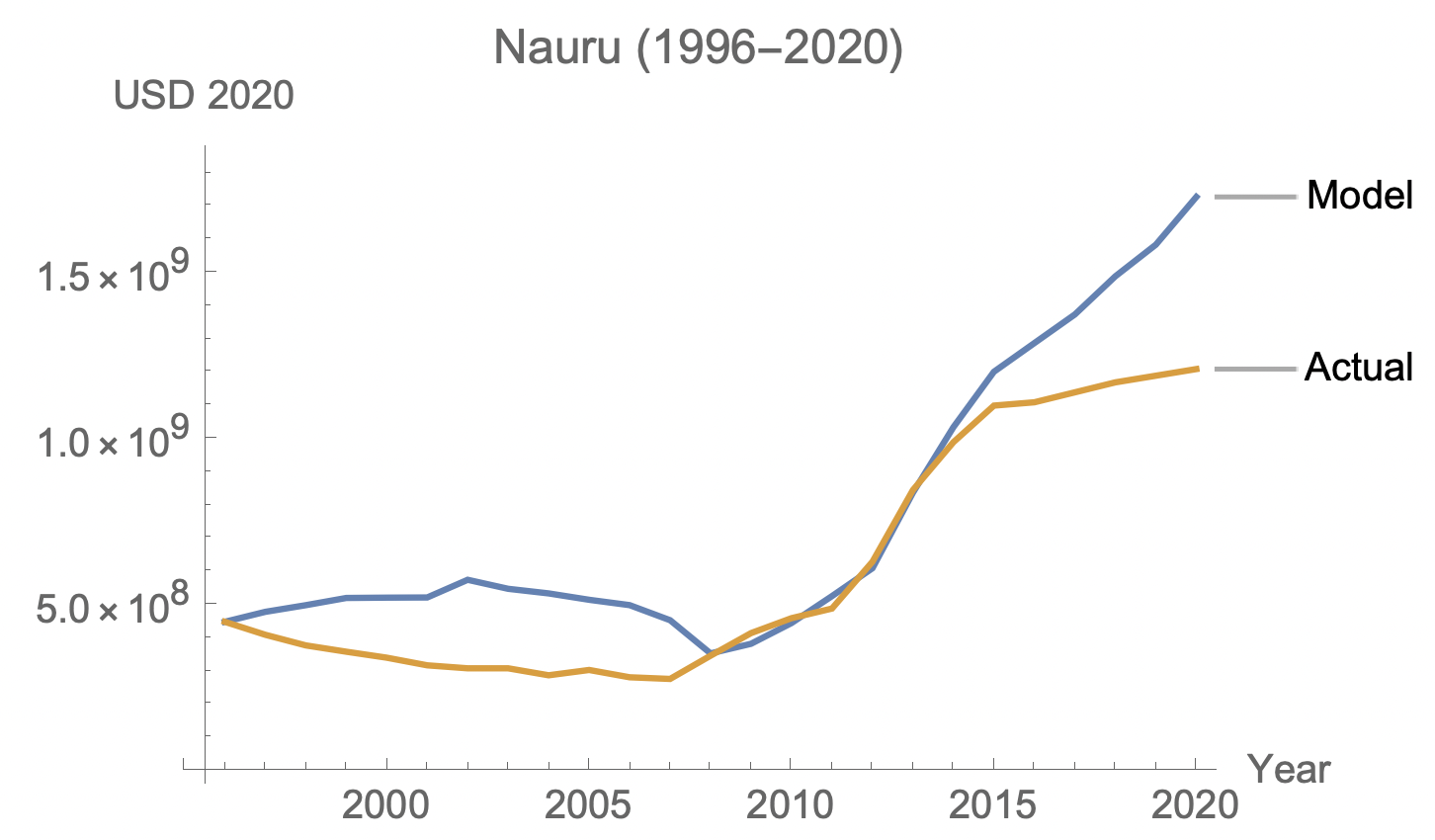}
\caption{Four examples of the model compared to actual national wealth.}
\end{figure}

The results are more dramatic for countries that experienced major violent conflict. Figure 7 shows four of these countries, with the model producing dips in wealth during periods of war. The degree of each downturn in the model should not be taken literally, as the conflict expenditure amounts  were highly speculative. Despite this, it should be clear that the model can express reductions in national power that occur due to interstate conflict.

\begin{figure}[H]
\centering
\includegraphics[scale=.25]{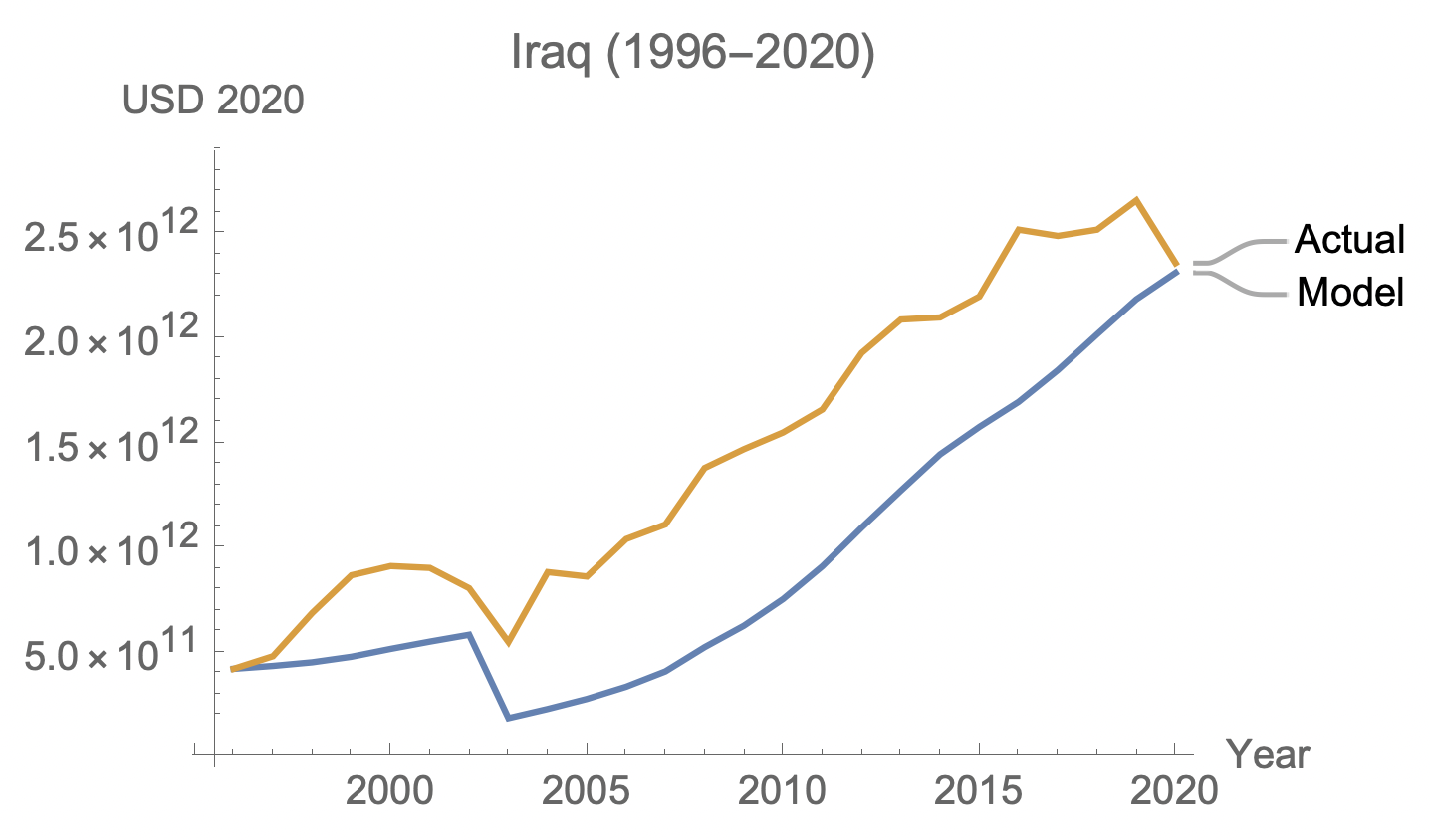}
\hspace{2em}
\includegraphics[scale=.25]{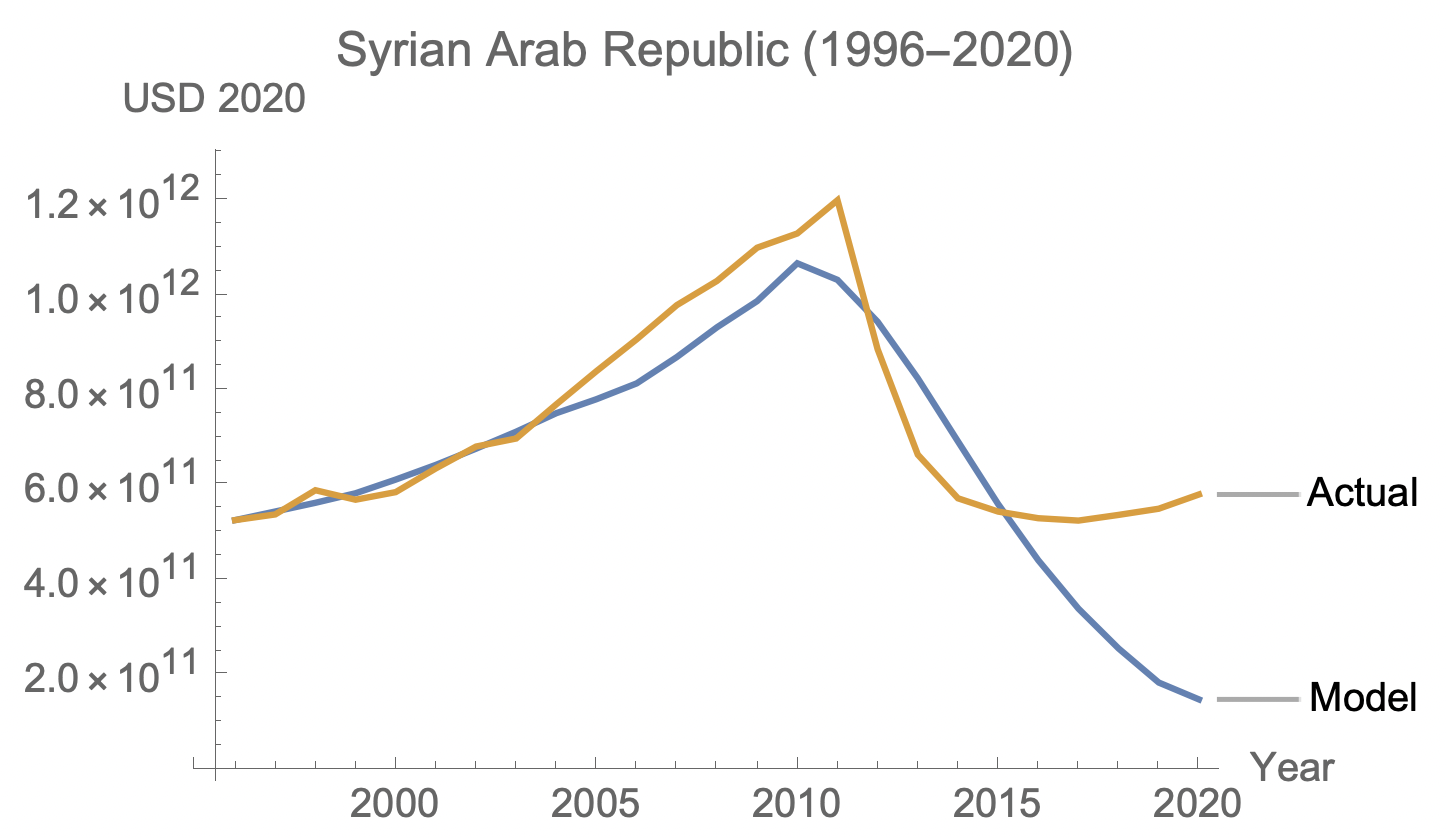}
\includegraphics[scale=.25]{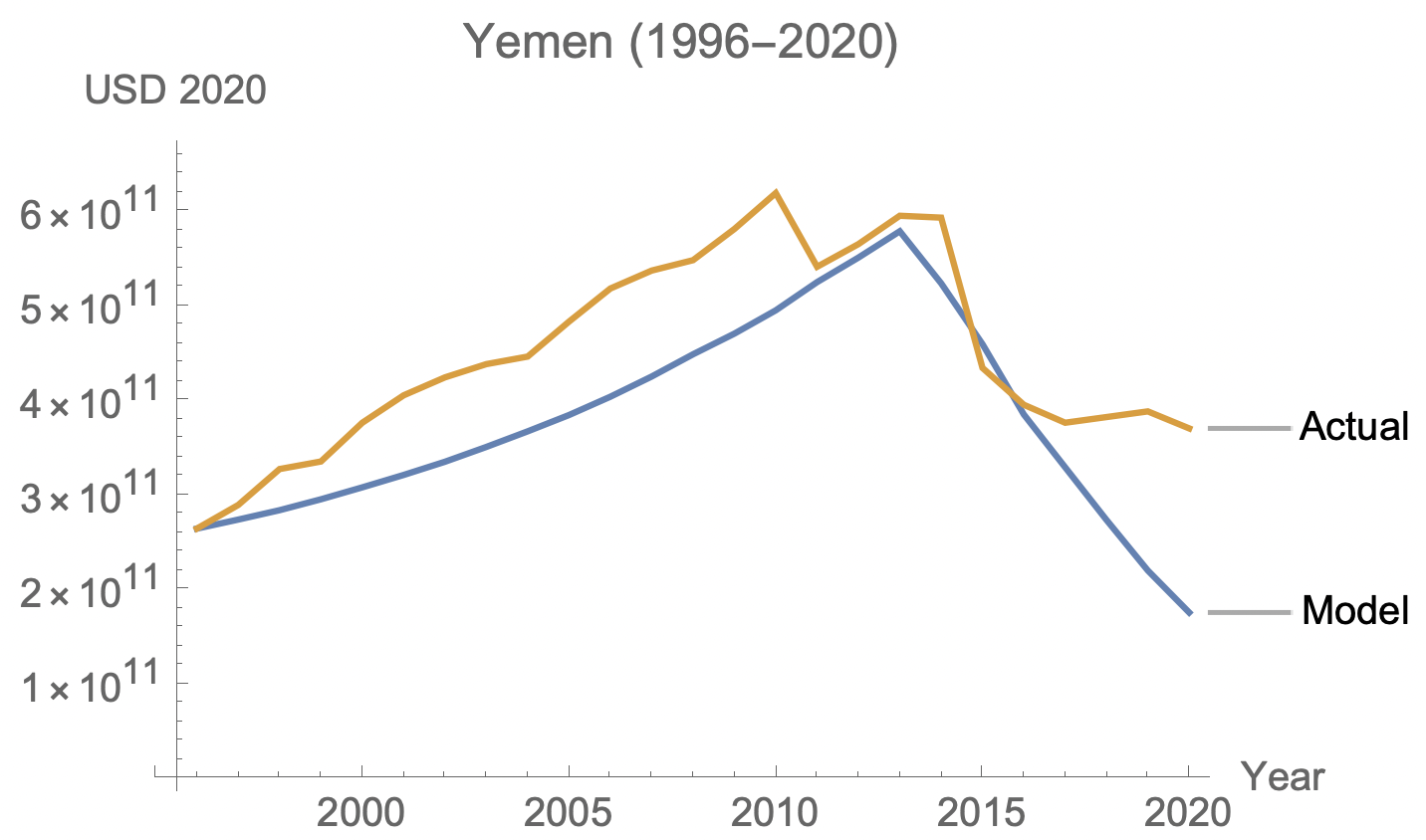}
\hspace{1em}
\includegraphics[scale=.25]{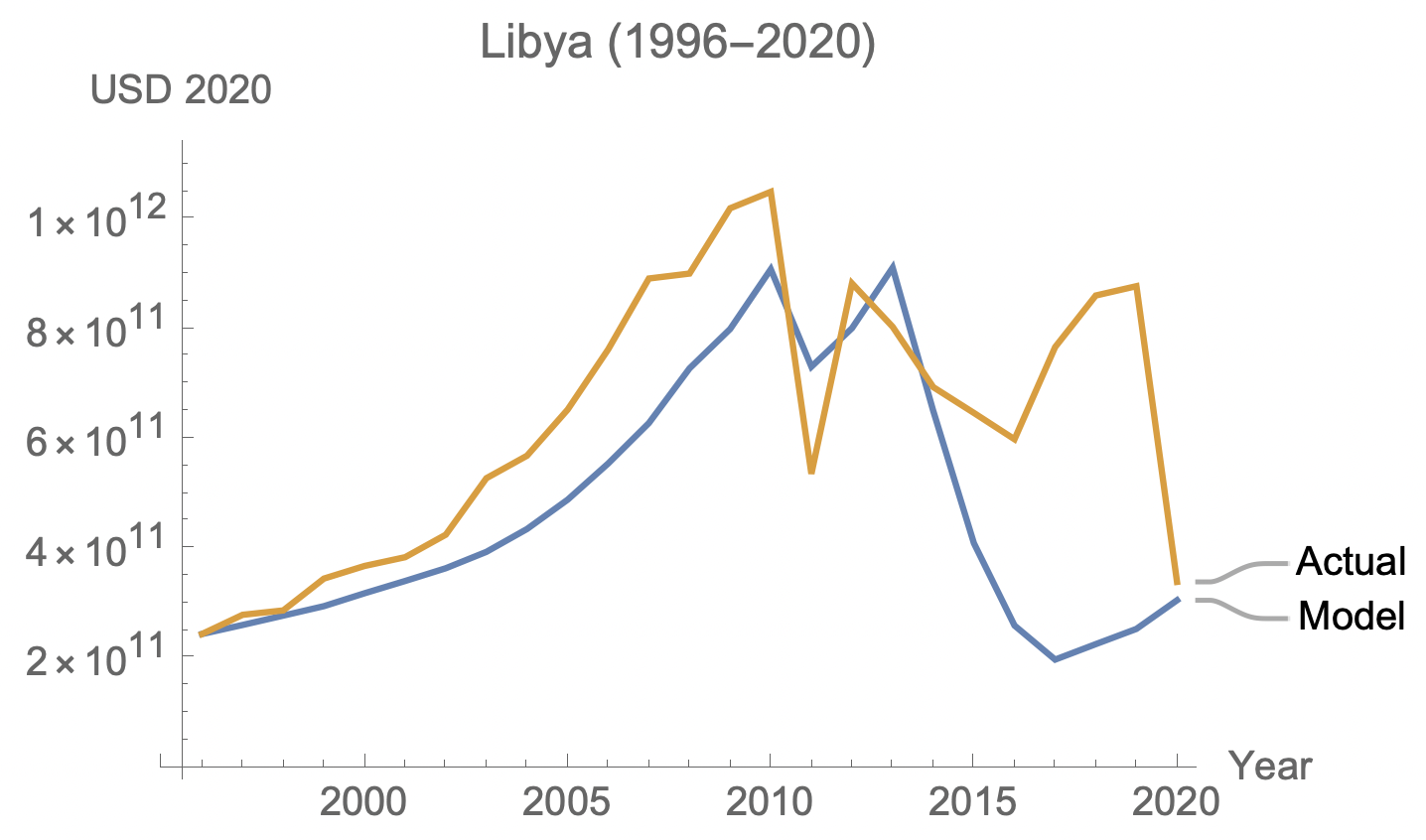}
\caption{Examples of countries that experienced major conflict.}
\end{figure}

The model can also be applied to historical periods outside the range of the baseline data set, assuming that the underlying data is available (or can be estimated) and that the parameter values established above remain valid. For example, in Figure 8 we illustrate the evolution of five major powers (Germany, Japan, Russia, the United Kingdom, and the United States) over the course of World War 2. Due to limited data availability, this simulation uses hypothetical values (see Appendix B). Nonetheless, it illustrates the reduction in strength of Germany and Japan as the war unfolded.

\begin{figure}[H]
\centering
\includegraphics[scale=.4]{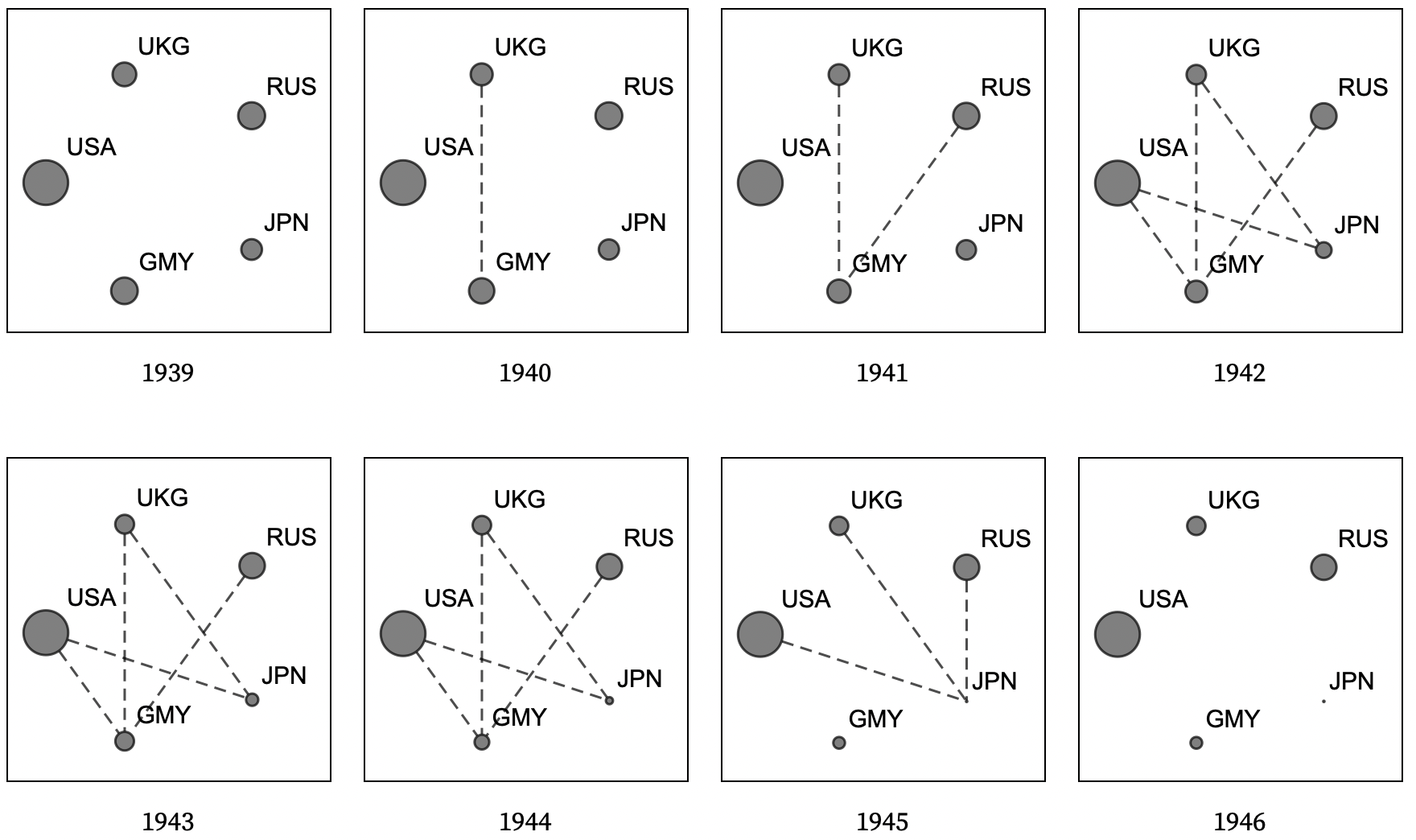}
\caption{Simplified power structure sequence depicting World War 2.}
\end{figure}

\subsection{Hypothetical Scenarios}

The power structure model can also be used to explore hypothetical situations. For example, one could ask how powerful Syria would be if it had not descended into civil war in 2011. Fig. 9 simulates that counterfactual scenario by assuming that the war never happened.

\begin{figure}[H]
\centering
\includegraphics[scale=.25]{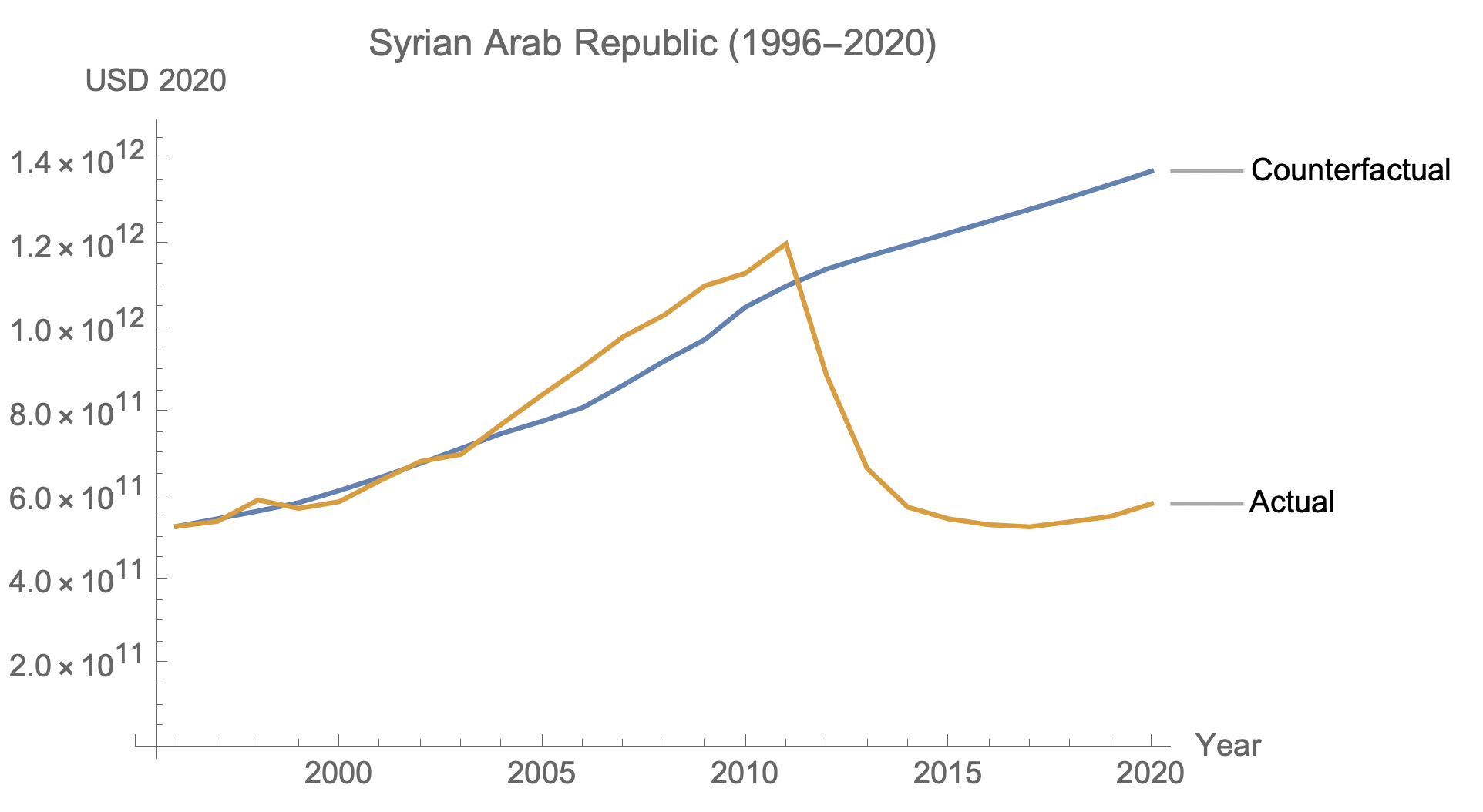}
\caption{Predicted wealth of Syria had it not experienced civil war starting in 2011.}
\end{figure}

Another example is Russia's invasion of Ukraine, which began six months ago as of current writing and which continues unabated. To model this war, the  major phenomena to be addressed are: (1) Russia is directing destructive power at Ukraine; (2) to a lesser extent, and defensively, Ukraine is directing destructive power at Russia; (3) third parties, such as the U.S. and Europe, are providing military and economic support to Ukraine; and (4) the U.S. and its allies have imposed economic sanctions on Russia. An additional consideration, which do not model (but could), is the massive transfer of wealth that would occur if Russia were to conquer and annex a portion of Ukrainian territory.

Even though this war began in February 2022, we use national wealth data from 2020, the last year of our data set. From this starting point, we project the strength of the two combatants after one year of war. We assume that Russia expends \$100B on the war, Ukraine spends \$25B, allies give Ukraine \$20B in assistance, Russian trade decreases by 20\%, and Ukraine's trade is cut in half. Plugging these estimates into the law of motion, the model predicts that Ukraine's strength will reduce to about 40\% of its prewar level, and that Russia's will remain about the same (the effects of the war and sanctions are offset by its intrinsic growth). This is not to suggest that Russia is unharmed by the conflict; its wealth would have increased in the absence of it.

Numerous other hypothetical examples such as these could be explored, provided that one can make educated guesses about changes in wealth levels, the effects of trade agreements and embargoes, and military expenditure. The model provides a framework for making  such back-of-the-envelope calculations about the evolution of power in the international system.

\subsection{The World Power Structure}

A power structure consists of actors of varying capacities, along with their relationships. By combining national wealth, trade, and military data as approximations that feed into the law of motion, we have modeled the \textit{world power structure}. This is a simplification of the world as it really is: The variables we used are mere proxies for the power relationships that we are trying to quantify. Moreover, we have only included nation states, thereby glossing over a much larger, more nuanced, and conflict-ridden network of actors.

\begin{figure}[H]
\centering
\includegraphics[scale=.15]{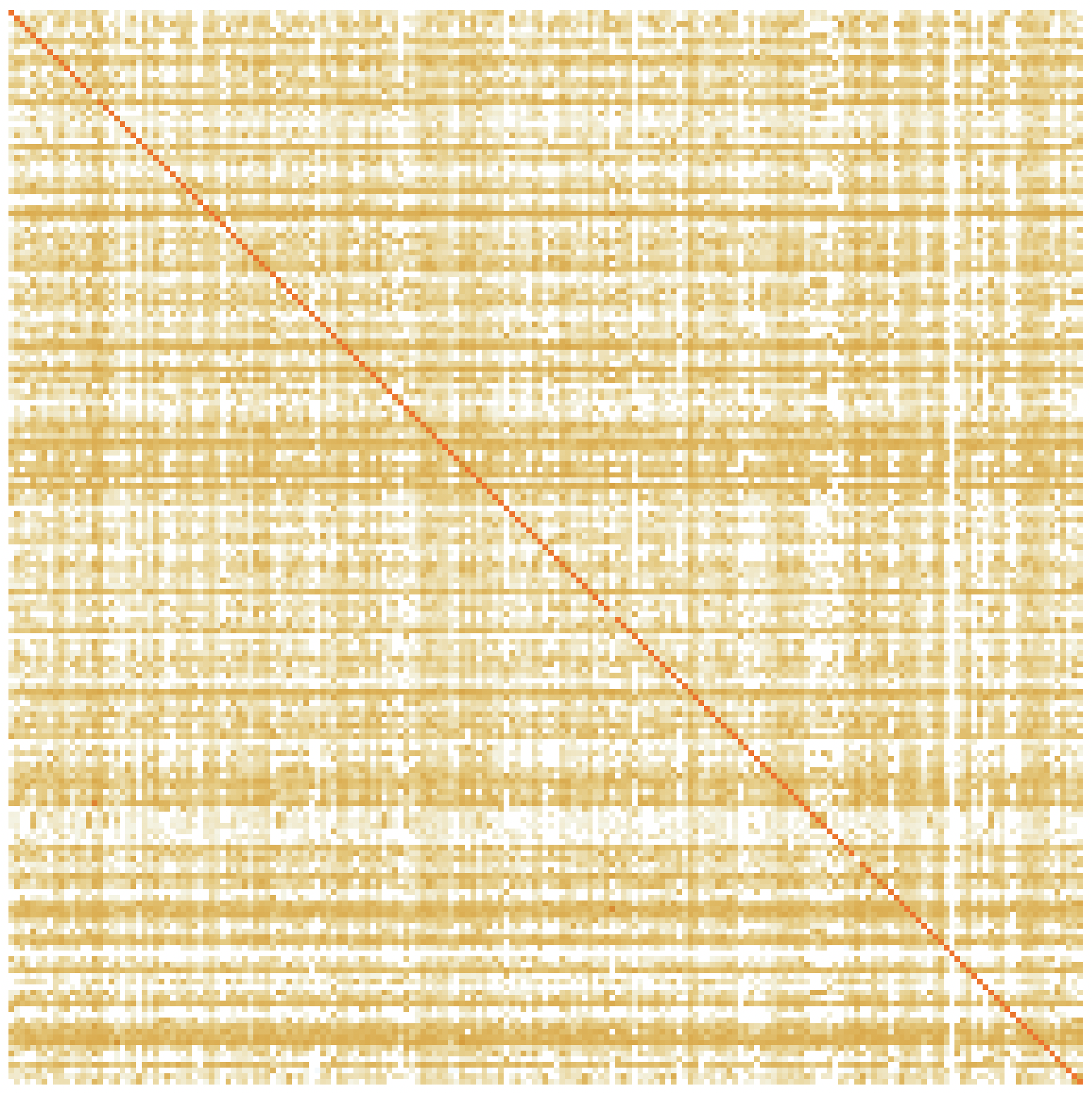}
\caption{World power structure: interstate relationships in 2020.}
\end{figure}

The interstate relationships of the world power structure as of 2020 are depicted in Figure 10. Each column in this 193x193 matrix represents a state's ``foreign policy" towards every other state, with brighter colors denoting more activity. Negative values, indicating destruction, are in blue, but there are only a few dozen of them and they are so faint as to be invisible, illustrating that the vast majority of state interaction in the contemporary international system is constructive. The orange diagonal is the power that each state does not set in motion but retains as a stock.

Even though this model of the world power structure is a simplification, we can distill an even simpler structure within it by plotting it as a graph in which each state is connected only to its primary trading partner (Fig. 11). 

\begin{figure}[H]
\centering
\includegraphics[scale=.24]{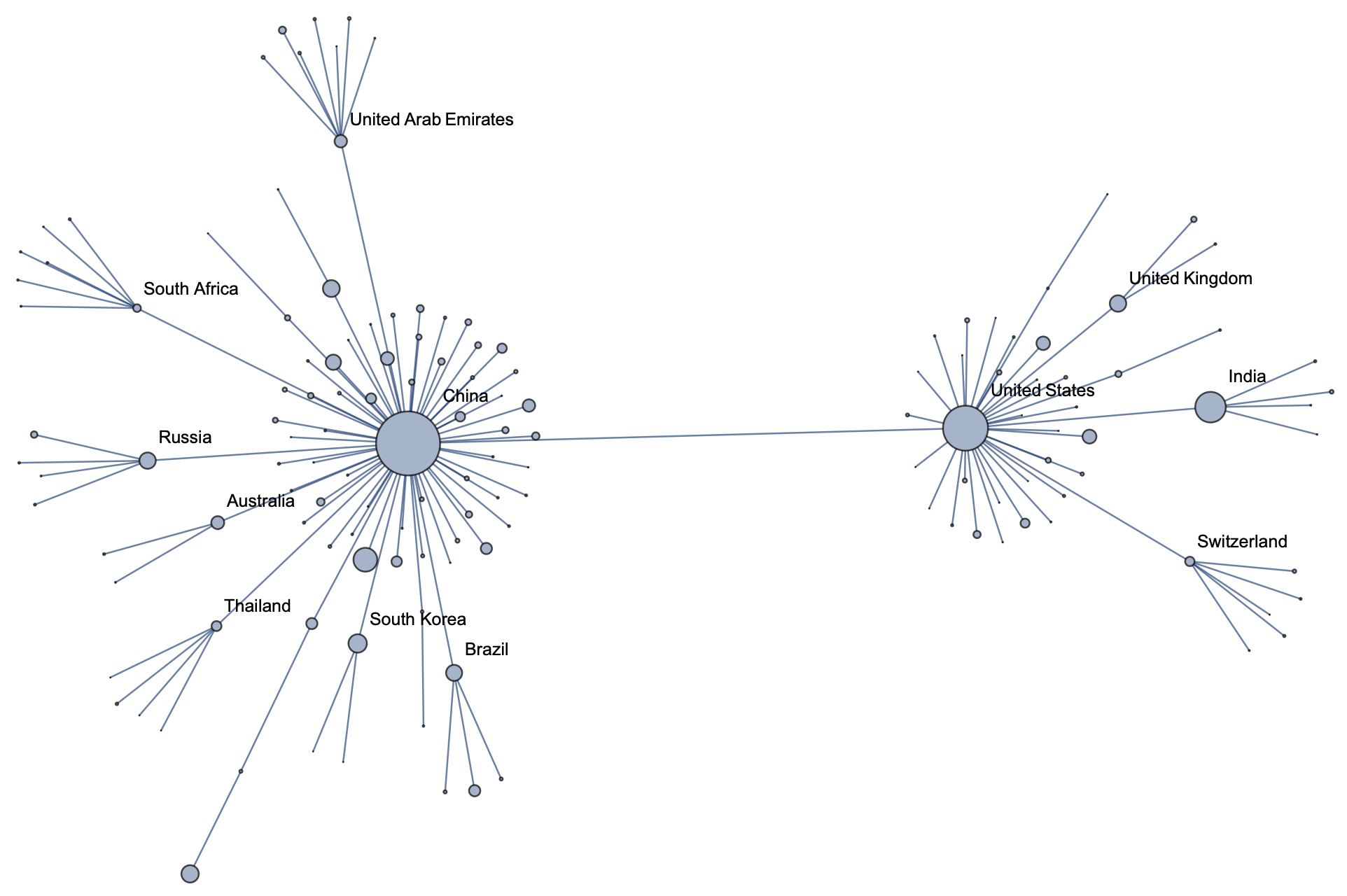}
\includegraphics[scale=.26]{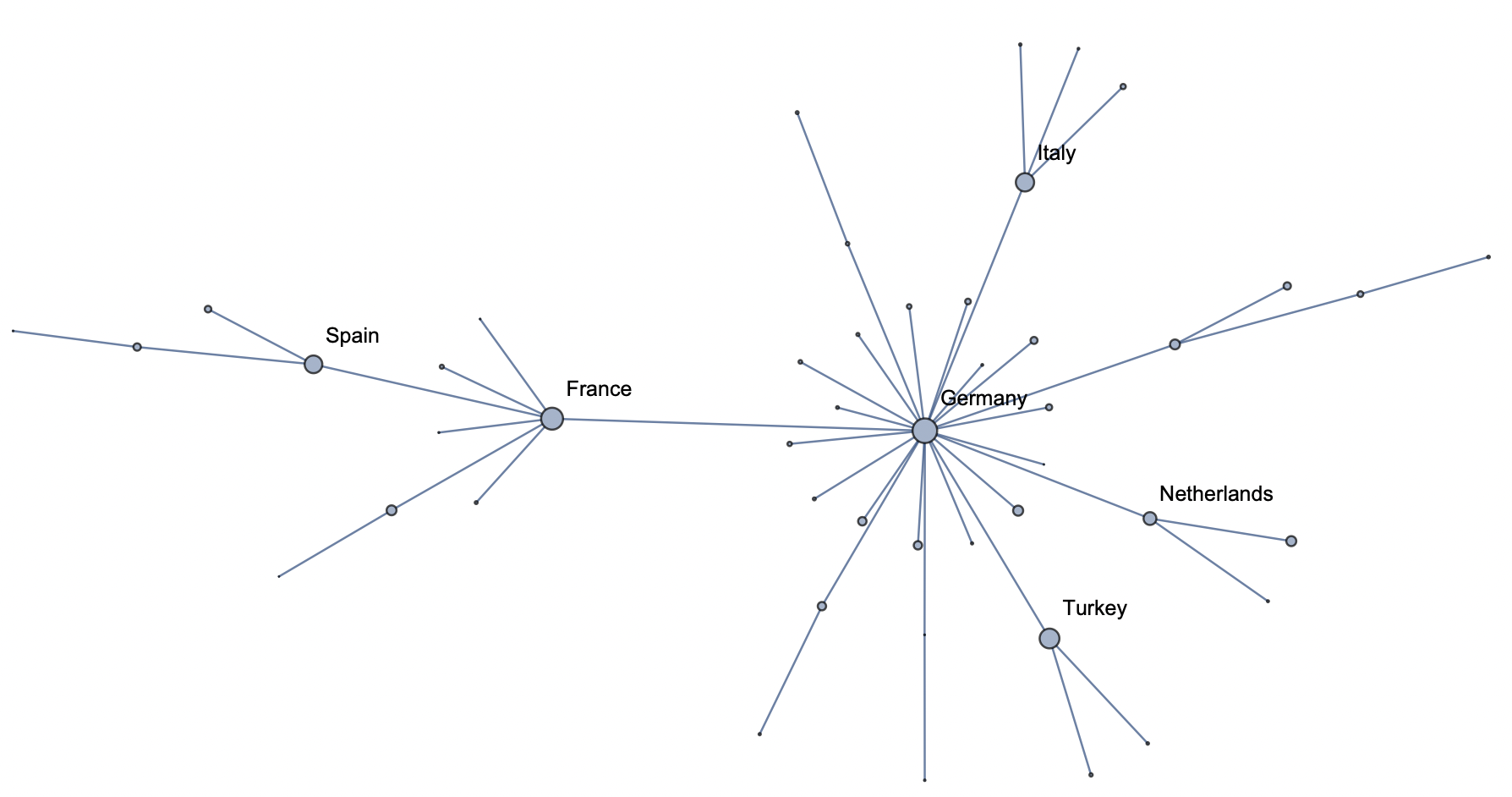}
\caption{World power structure: primary trading partner relationships, 2020.}
\end{figure}

When we run a naive simulation of the law of motion based on the 2020 world power structure, the model predicts an increasingly unipolar system dominated by China (Fig. 12).
\begin{figure}[H]
\centering
\includegraphics[scale=.3]{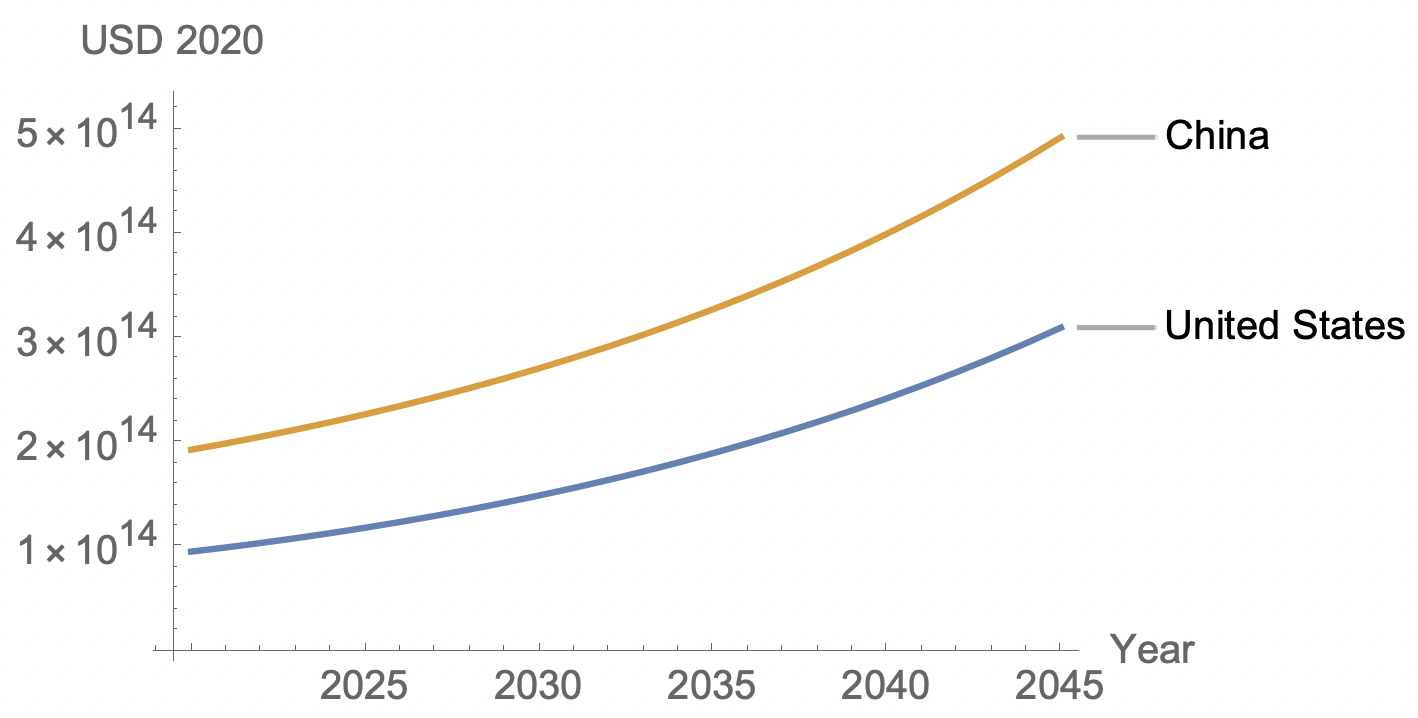}
\caption{Projected growth in power, China and the U.S.}
\end{figure}
\noindent
This unipolarity also exists when we model coalitions that might emerge around each of the two great powers. In (Poulshock 2022), we tried to forecast the development of these coalitions based solely upon economic incentives. Fig. 13 uses a representative scenario from that paper to show how the power of these two coalitions is likely evolve (left). It also shows a scenario in which a U.S.-led coalition forms to establish a balance of power against China (right).\footnote{In this simulation, members of China's coalition are Indonesia, Iran, Nigeria, Pakistan, Russia, and the United Arab Emirates. Members of the U.S.'s balancing coalition are Australia, Canada, France, Germany, Japan, South Korea, Spain, Taiwan, and the United Kingdom.} In both cases, the model naively simulates the future, starting in 2020, based on assumed closer trade ties among the members of each coalition.\footnote{This simulation assumes that members of each coalition reallocate 10\% of their trade away from members of the opposing coalition and toward members of their own coalition.}

\begin{figure}[H]
\centering
\includegraphics[scale=.3]{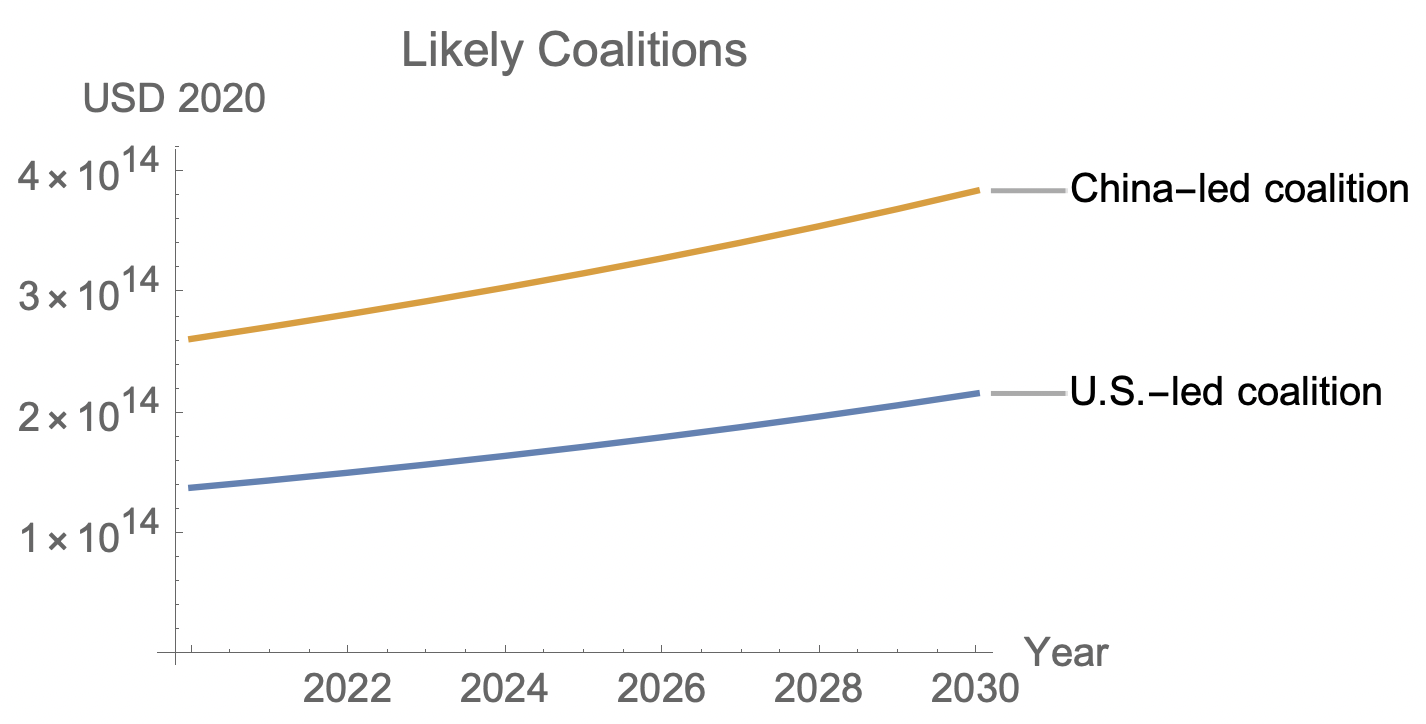}
\hspace{2em}
\includegraphics[scale=.3]{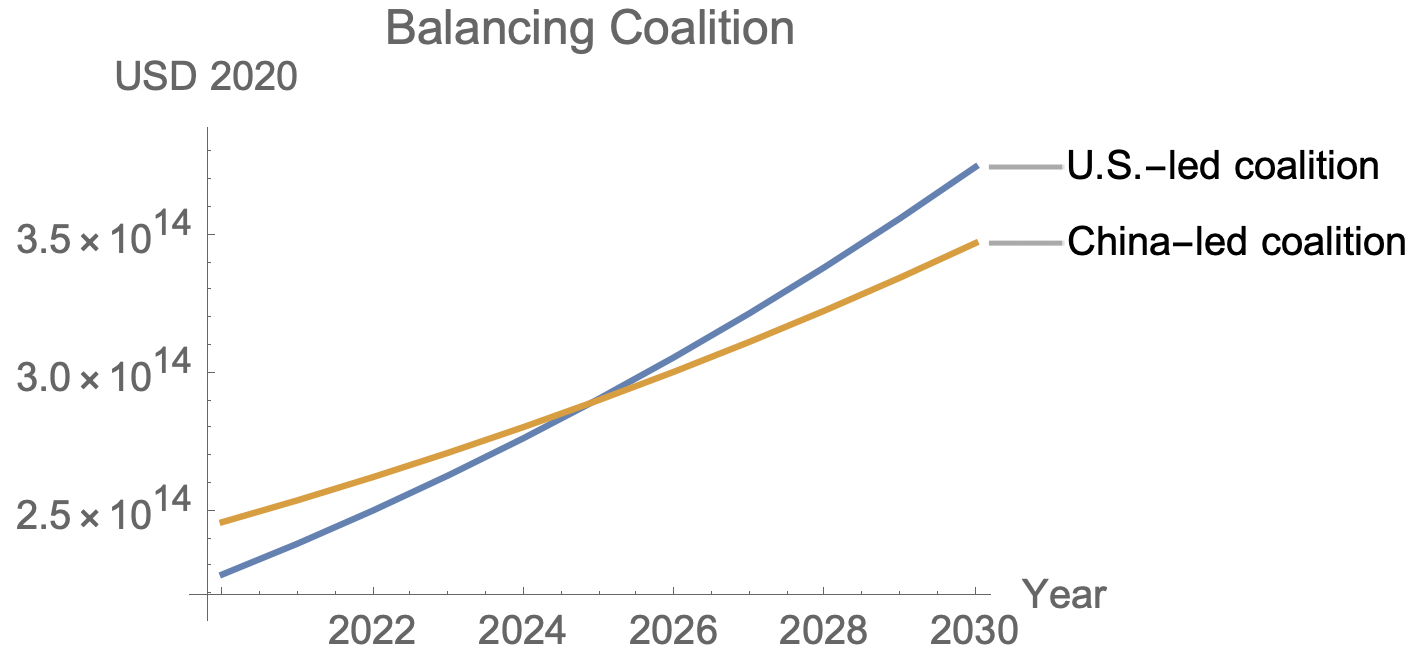}
\caption{Projected power of coalitions led by the U.S. and China: likely (left) and a U.S.-led balancing coalition (right).}
\end{figure}

This model of the world power structure can be used to predict the evolution of power in the international system and, as just illustrated, to explore hypothetical configurations of states.

\section{Conclusion}

We have proposed a way to quantify national power using wealth as a proxy variable. Constructive uses of wealth, such as trade, and destructive ones, such as military force, represent flows of power within the network of nation states. We used wealth, trade, and military data from 1995-2020 to parameterize a simple mathematical model that describes this flow of power. We then illustrated how the model can be used to simulate the evolution of power in actual and hypothetical scenarios.

We now reiterate the major caveats expressed above. First, there are many exogenous variables that affect national wealth beyond the ones used in this model, and these introduce a significant amount of variability into the data. The model attempts to reveal a general pattern within this noise. Second, wealth and power are not equivalent. The model is not meant to predict national wealth; it predicts power as roughly approximated by national wealth. Power suggests destructive capacity, and a more robust model would be able to distinguish between assets capable of constructive action and those capable of destruction. Finally, this model does not attempt to predict state behavior, such as what foreign policies countries will or should adopt. Instead, it quantifies the consequences of their behavior, that is, how power levels change as a result of trade and military policies.

The significance of this work is twofold. First, it provides a coherent definition of national power that is measurable in terms of real world quantities. Second, it offers a way to predict the outcomes of state interactions, and as a result it could potentially be used to explore the consequences of hypothetical state choices.

Future work includes the acquisition of more fine-grained data, particularly related to military assets and operations. It would also extend the world power structure data set to include non-state actors. 


\section{References}

\begin{footnotesize}
\begin{hangparas}{.2in}{1}

Correlates of War Project, \textit{Militarized Interstate Disputes Data Set (v4.3) (1816-2010)}.

Daggett, Stephen. 2010. \textit{Costs of Major U.S. Wars.} Congressional Research Service. Report 7-5700, RS22926.

Mearsheimer, John J. 2014. \textit{The Tragedy of Great Power Politics}. W. W. Norton \& Company.

Mulder, Nicholas. 2022. \textit{The Economic Weapon: The Rise of Sanctions as a Tool of Modern War}. Yale University Press.

Piketty, Thomas. 2014. \textit{Capital in the Twenty-First Century}. Harvard University Press.

Poulshock, Michael. 2019. \textit{The Foundations of Political Realism}. arXiv:1910.04785.

Poulshock, Michael. 2017. \textit{An Abstract Model of Historical Processes}. Cliodynamics: The Journal of Quantitative History and Cultural Evolution, issue 8(1).

Singer, J. David, Stuart Bremer, and John Stuckey. 1972.  ``Capability Distribution, Uncertainty, and Major Power War, 1820-1965." in Bruce Russett (ed), \textit{Peace, War, and Numbers}, Beverly Hills: Sage, pp. 19-48.

Smith, Jason M. and Halgin, Daniel and Kidwell, Virginie and Labianca, Giuseppe (Joe) and Brass, Daniel and Borgatti, Steve. 2012. \textit{Power in Politically Charged Networks}. Available at SSRN: https://ssrn.com/ abstract=2239486.

Stockholm International Peace Research Institute (SIPRI). \textit{Military Expenditure Database}. Available at: www.sipri.org/databases (accessed Sept. 8, 2021).

Tellis, Ashley J., et al. 2000. \textit{Measuring National Power in the Postindustrial Age}. RAND Corporation.

United Nations. 2008. \textit{System of National Accounts}.

Wolfram Research, Inc. 2022. \textit{Mathematica}, Version 12, Champaign, IL.

World Inequality Lab. \textit{World Inequality Data Set}. Available at https://wid.world/data/ (accessed Sept. 4, 2021).

World Trade Organization. \textit{WTO Data Portal}. Available at https://data.wto.org/ (accessed Sept. 5, 2021).

\end{hangparas}
\end{footnotesize}

\pagebreak
\section{Appendix A: Mathematical Model}
\small 

A power structure is composed of a size vector and a tactic matrix, \{\textbf{s}, \textbf{T}\}. The size vector \textbf{s} lists the power or wealth levels of each country. The tactic matrix \textbf{T} is composed of three matrices, representing constructive, destructive, and self- allocations of power:

\begin{equation}
    \mathbf{T} = \mathbf{T}^{+} - \mathbf{T}^{-} + \mathbf{T}^{0}
\end{equation}
Within each matrix, column vectors represent the states' foreign policies, which are expressed as percentages of the total power allocated.

Axiom 1 says that a constructive action is a transfer of power from one state that increases the power of another state. If state A engages in a constructive action, of magnitude \textit{x}, with state B, then state A's power decreases by \textit{x} and state B's power increases by more than \textit{x}. The resulting changes to the states' power levels, or sizes \textit{s}, can be represented as
\begin{align}
\Delta s_{\scriptscriptstyle{A}} &= -x \\
\Delta s_{\scriptscriptstyle{B}} &= \beta x
\end{align}
where the parameter $\beta > 1$ quantifies the magnitude of agent B's increase. So if two states exchange $x$ amount of constructive power, they will both increase in size by $\beta x - x = x (\beta -1)$.

Axiom 2 says that a destructive action is a transfer of power from one state that decreases the power of another state. If state A acts destructively towards state B with a magnitude of \textit{x}, then state A's power decreases by \textit{x} and state B's power decreases by at least \textit{x}, or
\begin{align}
\Delta s_{\scriptscriptstyle{A}} &= -x \\
\Delta s_{\scriptscriptstyle{B}} &= -\mu x
\end{align}
where the parameter $\mu > \beta$ quantifies the magnitude of state B's decrease. If two states exchange $x$ amount of destructive power, they will both decrease in size by $\mu x + x = x(\mu+1)$.
 
Axiom 3 says that unused power depreciates or dissipates relative to power used constructively. In other words, states that trade grow more than states that don't. This \textit{intrinsic growth rate} is governed by the parameter $\lambda$. If $\beta$ and $\lambda$ vary by country and year, they are vectors.

The law of motion follows directly from the three axioms and describes how the size of each country changes as power is transferred according to the tactic matrices:

\begin{equation}
    \mathbf{s}(t+1) = \mathrm{Ramp}((\boldsymbol{\beta} \mathbf{T}^{+} - \mu \mathbf{T}^{-} + \boldsymbol{\lambda} \mathbf{T}^{0}) . \mathbf{s}(t))
\end{equation}
The Ramp function prevents states from having negative sizes.

A complete discussion of the model, its axiomatic basis, and its relationship to political realism can be found in (Poulshock 2019).

\pagebreak
\section{Appendix B: Model Application}
\small 

\subsection{Data Sources}

The following data sources were used, with all currency values converted to USD 2020:

\textbf{National wealth.} We relied on the World Inequality Database (WID) for market valuations of national wealth based on purchasing price parity. We used data from 193 countries, covering the years 1995-2020. Market value was used because book value statistics were not available for China. We chose not to use the World Bank's Changing Wealth of Nations 2018 data set because it includes estimates of human capital, which have not traditionally been part of the definition of national wealth. We also did not use the Credit Suisse national wealth data set (2010-2020), because it reflects household wealth only. Both of these other data sources also had less temporal coverage than the WID.

\textbf{International trade.} For dyadic international trade statistics, we used the following three sources: (1) IMF Direction of Trade Statistics (DOTS), \textit{Goods, Value of Imports, Cost, Insurance, Freight} (TMG{\_}CIF{\_}USD) (1995-2020); (2) OECD Balanced Trade in Services dataset (BaTIS), \textit{Total trade in services (S200), Balanced Value} (1995-2012); and (3) World Trade Organization, \textit{Commercial services exports by sector and partner – annual, BOP6 - S - Memo item: Total services} (2013-2020). In future work, this trade data could be supplemented with primary and secondary income payments, which encompass investment income, compensation, and transfers. It could also be supplemented by unilateral transfers such as foreign aid, which we ignored as negligible relative to trade. For example, total U.S. foreign assistance in 2019 was approximately \$47B (see https://foreignassistance.gov/), or less than 1\% of the U.S.'s trade volume. We also ignored international financial flows, such as foreign investment, as these values can be negative and therefore problematic to interpret in this context.

\textbf{Military expenditure.} There are no available data sets covering the amount spent by individual countries on specific military conflicts. In lieu of that, we used annual military expenditure as a reasonable proxy variable setting an upper bound on the amount of destructive power that a state can allocate in a given year. We relied upon the Stockholm International Peace Research Institute's (SIPRI) Military Expenditure Database, variable \textit{Military Expenditure by Country}.

\textbf{Military conflict.} To estimate the amount of military resources expended on major interstate conflicts, we used Wikipedia to identify conflicts from 1995-2020 which averaged more than 25,000 deaths per year: the Afghan Civil War, Angolan Civil War, Eritrean-Ethiopian War, First and Second Congo Wars, Iraq War, Second Sudanese Civil War, Syrian Civil War, U.S. invasion of Afghanistan, and Yemeni Civil War. Intrastate conflicts were not included unless they served as proxy wars for third party states, in which case the third parties were included as aggressors. For each year and state dyad involved in these conflicts, we made order of magnitude estimates of the military expenditure or support. These numbers are highly speculative.

\textbf{World War 2.} This simplified sequence is based on back-of-the-napkin estimates. National wealth values were not available for Russia and Japan; we estimated these using cross-ratios to GDP. Applied destructive power was assumed to be 1\% of GDP, allocated per the Correlates of War Militarized Interstate Disputes data set. We ignored the effects of trade.

\subsection{Parameter Estimation}

\subsubsection{Effect of Destructive Action ($\mu$)}

To determine the value of $\mu$, we would ideally know combatants' expenditures on particular violent conflicts (specifically, their asset reductions) and the quantified damage that they inflicted upon their enemies. Because this data is not readily available, we approximate it by analyzing civil wars, which entail a state effectively using its military expenditure to destroy itself. We identified 10 states that experienced major internal conflicts between 1995 and 2020: Burundi, Cameroon, Cote d'Ivoire, Guinea-Bissau, Libya, Mali, Nepal, Serbia, Sierra Leone, and Yemen. We omitted other states where the necessary data was unavailable or where other variables would have confounded our estimate of peacetime growth. For each state, we determined the average growth in national wealth during peacetime and then calculated the loss in each year of civil war by subtracting the actual wealth from the expected wealth, i.e. what the national wealth would likely have been in the absence of war. 

In an internal conflict, the loss $l$ of national wealth is due to both the military power expended $x$ and the damage inflicted by that power. In other words 
\begin{equation}
     l = x + \mu x
\end{equation}
Solving for $\mu$
\begin{equation}
     \mu = \frac{l-x}{x}
\end{equation}
We then calculated $\mu$ for 40 country-years for which there was data, selecting conflicts severe enough that $x$ was likely in the realm of annual military expenditure. The average value for $\mu$ was 27.4, which we round to 30 due to the omission of some high outliers and to convey the large uncertainty of this estimate. Though we used a very small sample and an imperfect methodology, a destruction factor of 30x does not seem unreasonable for modern violent conflict. Given the low incidence of interstate war, it will not alter the simulation results significantly even if this back-of-the-napkin approximation is off by a factor of 5.

\subsubsection{Growth Rates ($\beta$, $\lambda$)}

To estimate the intrinsic growth rate ($\lambda$) and the growth rate due to trade ($\beta$), we first analyzed the annual growth in national wealth as a function of trade percentage (that is, trade volume divided by national wealth). Though the data is widely dispersed (see Fig. 6), this yields a linear relationship of
\begin{equation}
    g = 1.025 + 0.201 \, tp
\end{equation}
where $g$ is the annual growth rate and $tp$ is the trade percentage. The first term in this equation suggests that the growth rate in the absence of trade, in other words, the intrinsic growth rate, is around 2.5\%. Note that while we applied this method globally to the entire data set, one could also apply it to individual countries as a way of accounting for differences in their intrinsic growth rates.

\begin{figure}[H]
\centering
\includegraphics[scale=.25]{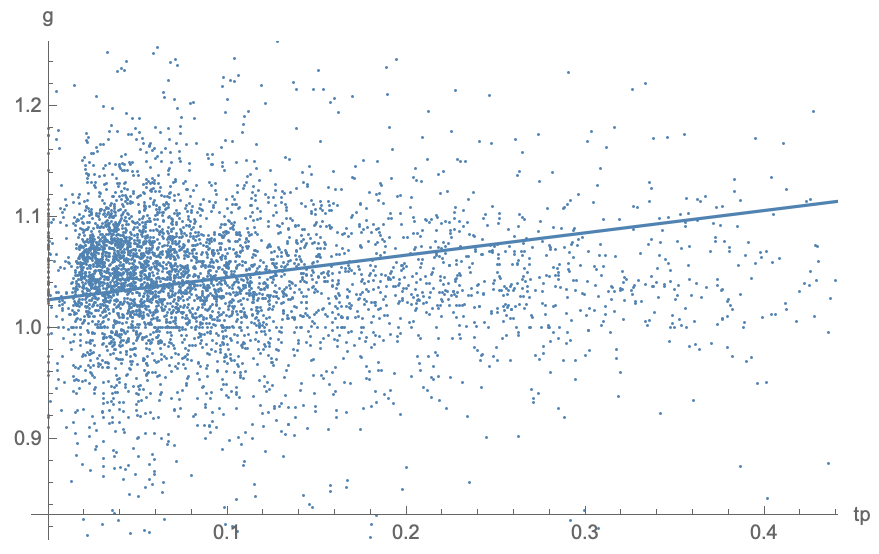}
\caption{Percentage annual growth in national wealth as a function of trade percentage.}
\end{figure}

Given the values of $\mu$ and $\lambda$ obtained above, we then estimated $\beta$ by simulating the law of motion in each year and finding the value of $\beta$ that minimized the average Euclidean distance between the actual wealth vectors and the predicted ones. This resulted in $\beta = 1.392$.

\end{document}